\documentclass[page-classic]{epl2}

\begin{document}

\title{Exchange bias with Fe substitution in LaMnO$_3$}

\author{M. Patra \and K. De \and S. Majumdar \and S. Giri\thanks{E-mail: \email{sspsg2@iacs.res.in}}}

\institute{Department of Solid State Physics, Indian Association for the Cultivation of Science, Jadavpur, Kolkata 700 032, India}


\abstract
{The exchange bias (EB) in LaMn$_{0.7}$Fe$_{0.3}$O$_3$ is observed by the negative shift and training effect of the hysteresis loops, while the sample was cooled in external magnetic field. The analysis of cooling field dependence of EB gives the size of the ferromagnetic (FM) cluster $\approx$ 25 $\AA$, where the magnetic anisotropy of FM cluster is found two order of magnitude higher than the FM bulk manganites. We propose that the nanoscale FM clusters are embedded in the glassy magnetic host with EB at the FM/glassy magnetic interface.}

\maketitle

\section{Introduction}
Recently, the observation of exchange bias (EB) in the phase separated (PS) charge ordered (CO) manganites opens up a new prospective of technological applications in addition to the colossal magnetoresistance (CMR) properties. ~\cite{salamon,qian} Exchange bias is a novel phenomenon, which is ascribed to the induced exchange anisotropy at the interface between ferromagnetic (FM) and antiferromagnetic (AFM) phases in a heterogeneous system. ~\cite{nogues, stamps} The induced exchange anisotropy is unidirectional in character and increases the effective anisotropy of the heterogeneous system, which has the technological applications for storage and spin-electron devices. The evidence of EB was first discovered by Meiklejohn and Bean in 1956 for FM Co core and AFM CoO shell structure, ~\cite{meik} which has also been observed at the FM/spin-glass (SG) interface. ~\cite{ali, gruyters, tang1, tang2}

\begin{figure}
\onefigure{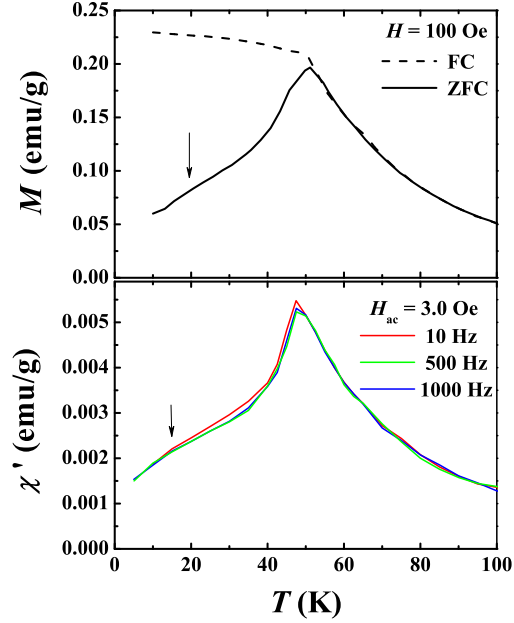}
\caption{(top panel) Temperature dependence of ZFC-FC magnetization and (bottom panel) $\chi^{\prime}$ with temperature at different frequencies.}
\label{fig.1}
\end{figure}

In addition to the technological importance, the evidence of EB further provides the microscopic views of the inhomogeneous phase separation in manganites. The first evidence of EB was observed in CO manganite Pr$_{1/3}$Ca$_{2/3}$MnO$_3$, where ferromagnetic (FM) droplets are naturally embedded in the AFM background. ~\cite{salamon} The other example of EB have recently been reported for CO manganite, where the strong cooling field dependence of EB is ascribed to the thickness of the FM layer in a spontaneous lamellar
ferromagnetic/antiferromagnetic phase separated Y$_{0.2}$Ca$_{0.8}$MnO$_3$. ~\cite{qian} In this letter, we also report the EB for 30 \% Fe substitution in LaMnO$_3$ at the FM/glassy magnetic interface. {\it Notably, we observe the EB at the FM/glassy magnetic interface in contrast to the phenomanon at the FM/AFM interface for CO manganites}. The EB was characterised by the unidirectional shift of the magnetic hysteresis loops under different field-cooled conditions, which was further confirmed by the training effect.

\section{Experimental procedure}
The polycrystalline sample of LaMn$_{0.7}$Fe$_{0.3}$O$_3$ was prepared by a chemical route. ~\cite{kde} The single rhombohedral phase ({\it R}$\overline{3}${\it c}) of the compound was characterized by powder x-ray diffraction pattern. The average size of the particle was $\sim$ 70 nm as observed using Transmission Electron Microscope. The ac and dc magnetizations were measured by a commercial superconducting quantum interference device (SQUID) magnetometer (MPMS, XL). In case of zero-field cooled measurements the sample was cooled down to the desired temperature from well above the transition temperature in zero magnetic field and measurements were performed in the heating cycle with magnetic field, while for field-cooled case the sample was cooled in presence of magnetic field and measured during heating the sample like zero-field cooled case. 

\section{Experimental results and discussions}
The temperature dependence of zero-field cooled (ZFC) and field-cooled (FC) magnetizations are shown in the top panel of fig.~\ref{fig.1}, where the ZFC magnetization exhibits a sharp peak ($T_{p}$) at 51 K. ~\cite{kde} In consistent with the dc results the real part ($\chi^{\prime}$) of ac susceptibility ($\chi_{ac}$) also shows a peak as seen in the bottom panel of fig.~\ref{fig.1}, where a small frequency dependence of $\chi^{\prime}$ is noticed in between $\sim$ 50 and $\sim$ 15 K. Note that a shoulder ($T_f$) in the low temperature region is observed for dc and ac measurements in addition to the sharp peak, which is shown by the arrows in fig.~\ref{fig.1}. ~\cite{imaginary}   

\begin{figure}
\onefigure{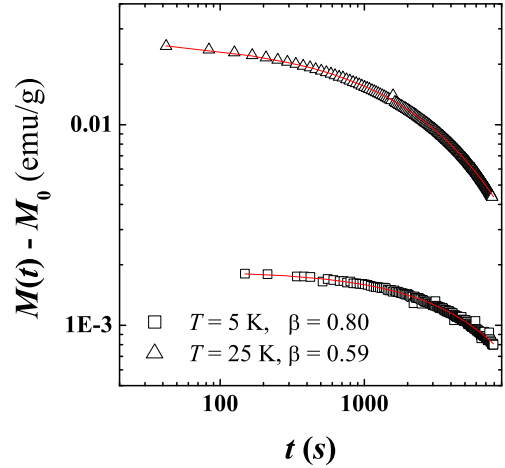}
\caption{Relaxation of remanent magnetization with $t$ at 5 K and 25 K .The solid lines exhibit the fit of the experimental data using stretched exponential.}
\label{fig.2}
\end{figure}

In order to understand the origin of weak frequency dependence of $\chi^{\prime}$ we measured dc magnetization with time ($t$). The sample was cooled down to 5 K and 25 K from 150 K ($\gg T_p$) with 100 Oe and then remanent magnetization was recorded with $t$ as seen in fig.~\ref{fig.2}. We considered the stretched exponential, $M(t) = M_0 - M _g\exp(-\frac{t}{\tau})^{\beta}$ to fit the relaxation of remanent magnetization.  
$M_0$ and $M_g$ are involved with the FM and glassy components, respectively. For typical SG compounds, the second term in the above expression was sufficient to fit the relaxation, where the values of $\beta$ are in the range of 0 $< \beta <$ 1. The term $M_0$ was incorporated to fit the relaxation for reentrant SG compounds. ~\cite{sinha} Here, $M_0$ was also required to fit the relaxation, while the values of $\beta$ are 0.80 and 0.59 for 5 and 25 K, respectively. The values of $\beta$ in addition to non-zero values of $M_0$ clearly indicates the signature of coexistence of FM and glassy magnetic components.

\begin{figure}
\onefigure{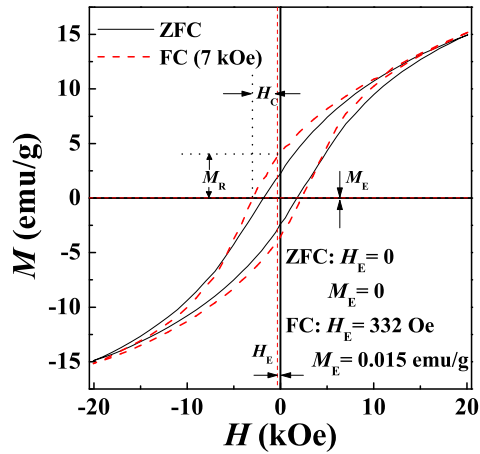}
\caption{Magnetic hysteresis loops measured at 5 K after field cooling with 7 kOe and zero field cooling. The details of the estimate of $H_E$, $M_E$, $H_C$, and $M_R$ are shown in the figure.}
\label{fig.3}
\end{figure}

\begin{figure}
\onefigure{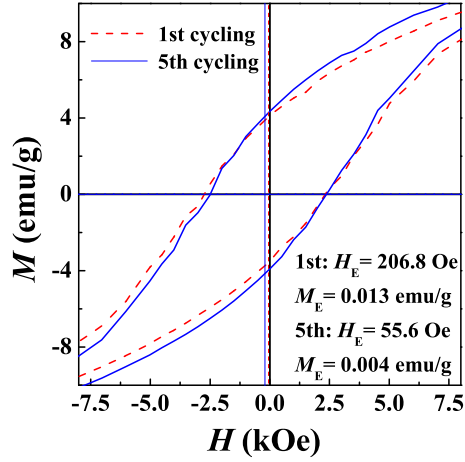}
\caption{Magnetic hysteresis loops after first and fifth field-cooled cycling with 10 kOe. The values of $H_E$ and $M_E$ after first and fifth cycling are given in the figure.}
\label{fig.4}
\end{figure}

The evidence of EB is observed by the the negative shift of the magnetic hysteresis loop, while the sample was cooled from 150 K ($\gg T_p$) to 5 K under FC condition with 7 kOe and then the hysteresis was measured in between $\pm$ 20.0 kOe at $5$ K after stabilizing the temperature. The shift is absent while cooling under ZFC condition as seen in fig.~\ref{fig.3}. The coercivity and the area of the hysteresis losses increase considerably in accordance with the common features of the EB. The exchange bias field ($H_E$), coercive field ($H_C$), EB magnetization ($M_E$), and remanent magnetization ($M_R$) were measured 332.0 Oe, 260.0 Oe, 0.0160 emu/g, and 3.79 emu/g, respectively from the gravity center of the shifted loop along field and magnetization axes as described in fig.~\ref{fig.3}. In order to confirm the EB we further studied the training effect, which describes the decrease of $H_E$, while several times thermal cycling the systems were performed. ~\cite{nogues, stamps} The hysteresis loops after first and fifth cycling are shown in fig.~\ref{fig.4}, where the decrease of the negative shift after five successive thermal cycling is noticed as seen in the figure. The vales of $H_E$ are 207 and 56 Oe, while for $M_E$ the values are 0.013 and 0.004 emu/g for first and fifth cycling, respectively. 

\begin{figure}
\onefigure{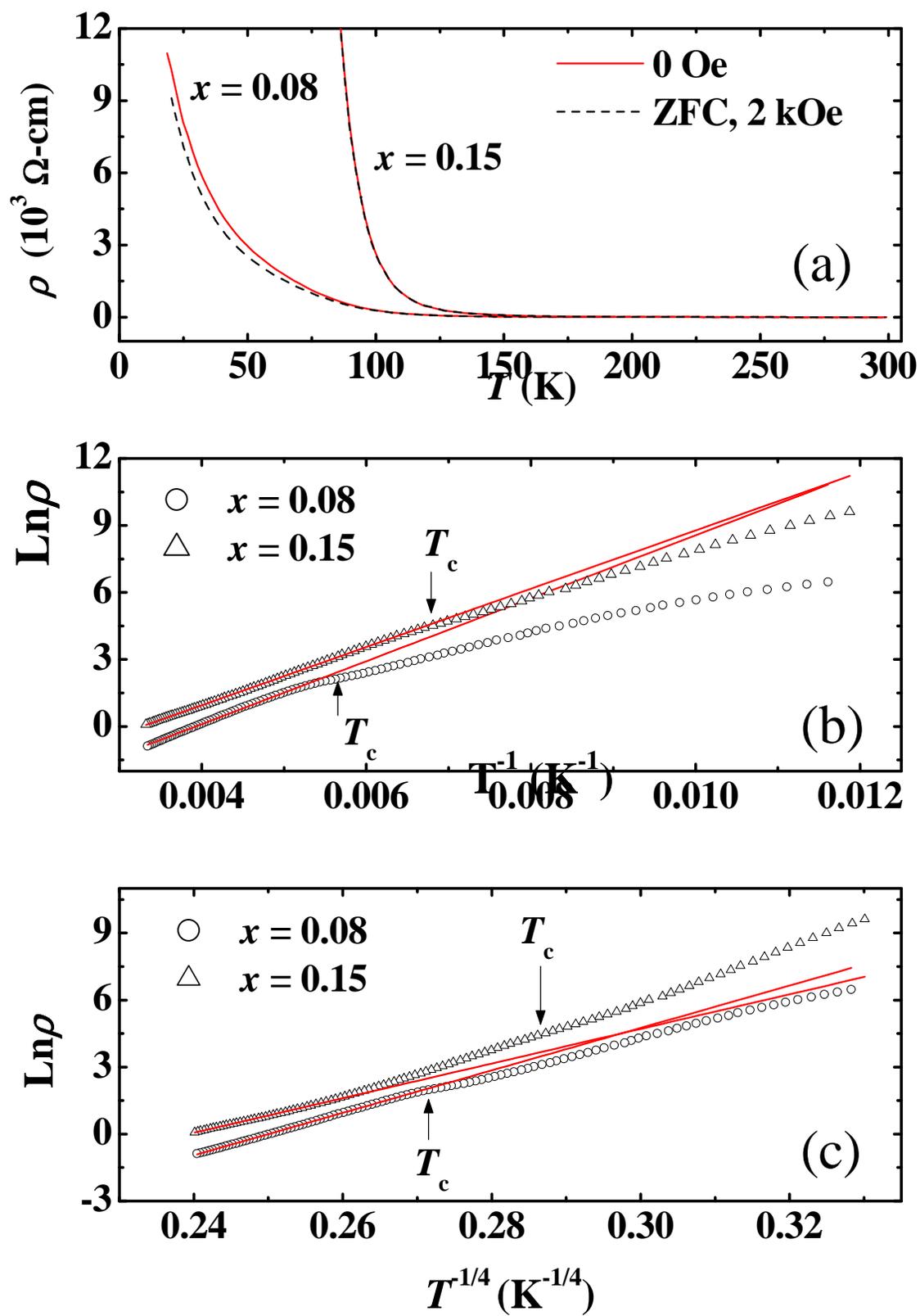}
\caption{Temperature dependence of $H_E$ and $M_E$. The arrows indicate $T_p$ and $T_f$.}
\label{fig.5}
\end{figure}


The temperature dependence of EB was studied, while the sample was cooled down to the desired temperatures from 150 K in 7 kOe field and the hysteresis loops were measured in between $\pm$ 20.0 kOe. The values of $H_E$ and $M_E$ as a function of temperature are shown in fig.~\ref{fig.5}. The $M_E$ decrease with increasing temperature, while the sharp decreasing trend is noticed below $\sim$ 50 K, which is close to $T_p$. Note that the EB is observed at the FM/glassy magnetic interface, where the glassy phase induces the 'frozen' FM spins at the interface resulting non-zero values of $H_E$. Here, $H_E$ decreases sharply with increasing temperature, which is almost zero above 20 K. Note that a shoulder ($T_f$) in the dc and ac magnetization is observed around $\sim$ 20 K, which indicates that $T_f$ exhibits the spin freezing temperature.  

\begin{figure}
\onefigure{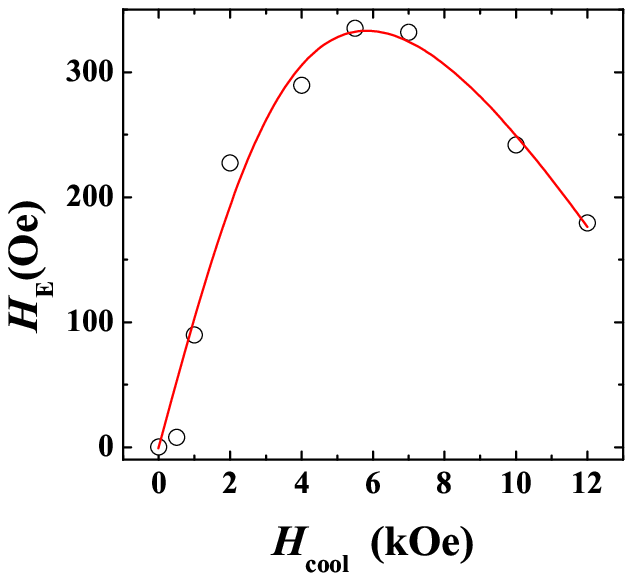}
\caption{The plot of $H_E$ as a function of $H_{cool}$. The continuous curve indicates the fit using equation described in the text.}
\label{fig.6}
\end{figure}
 

The EB was further studied at the different cooling field ($H_{cool}$), while the sample was cooled down to 5 K from 150 K in different $H_{cool}$ and then hysteresis was measured in between $\pm$ 20.0 kOe. The plot of $H_E$ as a function of $H_{cool}$ is shown in fig.~\ref{fig.6}, which increases with $H_{cool}$ and then decreases below $\sim$ 7 kOe with the further increase of $H_{cool}$. In order to understand $H_{cool}$ dependence of $H_E$ we used the following expression as, ~\cite{salamon} 
$H_E \propto J_i[\frac {J_i\mu_0}{(g \mu_B)^2}L(\frac {\mu H_{cool}}{k_BT_f})+H_{cool}]$ with approximation for $\mu H_{cool} < k_BT_f$, 
where $J_i$ is the interface exchange constant. For small FM clusters we assume the magnetic moment of FM cluster $\mu$ = $N_v\mu_0$  with $\mu_0$ $\approx$ 3$\mu_B$ for FM Mn core spin, where $N_v$ is number of spins per FM cluster. Note that  
the first term in the expression dominates for small $H_{cool}$ ($\propto j_i^2$), while for large $H_{cool}$ the second term ($\propto j_i$) dominates. The solid line in fig.~\ref{fig.6} exhibits the satisfactory fit of the experimental data using the above expression with $J_i$ and $\mu$ as adjustable parameters. The values of $J_i$ and $\mu$ are $\approx -0.143$ meV and 141$\mu_B$, respectively. Using the values of $\mu$, $N_v$ is $\approx$ 47 and the number density of FM clusters ($n$) can be estimated from the saturation magnetization as $M_s$ = $n\mu$, which gives $n \approx$ 11.25 $\times 10^{-5}$ $\AA^{-3}$. The value of $n$ further gives us the size of FM cluster $\approx$ 25 $\AA$, which is in consistent with the size of the FM droplets in inhomogeneously PS manganites based on the small angle neutron scattering. ~\cite{hennion, hennion1, gran} The value of magnetic anisotropy constant $K$ is estimated $\sim$ 0.3 $\times$ 10$^{7}$ erg/cm$^3$ at 5 K from the following expression as $M_E/M_S$ $\sim$ - 2$\nu_0\tau$ exp(-$KV/k_{B}T$) sinh($\mu H_E/k_{B}T$), where $V$ is the volume of the FM cluster. ~\cite{salamon} The values of typical measurement time ($\tau$) and the switching frequency of magnetization ($\nu_0$) are typically taken 10$^{3}$ s and 10$^9$ s$^{-1}$, respectively. Note that the value of $K$ is two order higher than  bulk FM manganites. ~\cite{suz, buk} The large anisotropy is ascribed to the induced exchange anisotropy at the FM/glassy magnetic interface, where exchange anisotropy is strongly dependent on the area of the interface. Here, we propose that the FM clusters are embedded in the glassy magnetic host in order to satisfy the large interface. The above scenario is analogous to the phase separation scenario proposed for low hole doped and CO manganites, where FM droplets are embedded in the AFM matrix. ~\cite{salamon, hennion, hennion1, gran} Recently, the features of cluster-glass state has been proposed for LaMn$_{0.7}$Fe$_{0.3}$O$_3$, where the glassy magnetic behavior is attributed to the competition between FM and AFM interactions in addition to the random substitution of Fe in Mn sites because of the same ionic radii of Fe$^{3+}$ and Mn$^{3+}$. ~\cite{kde, liu} Here, the evidence of EB further provides microscopic views of cluster-glass scenario. 

\section{Conclusion}
In conclusion, we observe the EB at the FM/glassy magnetic interface from the negative shift and training effect of the hysteresis loops after cooling the sample in external magnetic field. The magnetic anisotropy of the FM cluster is found two order of magnetic higher than the bulk FM manganites. We propose an interesting phase separation scenario of cluster-glass state in LaMn$_{0.7}$Fe$_{0.3}$O$_3$, where the nanoscale FM clusters are embedded in the glassy magnetic host. 

\acknowledgments
One of the authors (S.G.) wishes to thank CSIR, India for the financial support. The magnetization data were measured under the scheme of Unit on Nanoscience and Nanotechnology, IACS, Kolkata, India.


\begin{thebibliography}{99}
\bibitem{imaginary} Imaginary part of $\chi_{ac}$ was below SQUID resolution.
\end{thebibliography}
\end{document}